\documentclass[]{aa}
\usepackage[varg]{txfonts}
\usepackage{amsmath}
\usepackage{graphicx}
\usepackage{multirow}
\usepackage{url}

\begin{document}

\title{Comparative study of gamma-ray emission from molecular clouds and star-forming galaxies}
\titlerunning{Statistic study of gamma-ray emission}
\author{Fang-Kun Peng\inst{1,2,3,4}
\and Shao-Qiang Xi\inst{2,4}
\and Xiang-Yu Wang\inst{2,4}
\and Qi-Jun Zhi\inst{1}
\and Di Li\inst{5,6}}
\institute{Guizhou Provincial Key Laboratory of Radio Astronomy and Data Processing, Guizhou Normal University, Guiyang 550001, China\\ \email{pengfangkun@163.com}
\and School of Astronomy and Space Science, Nanjing University, Nanjing 210093, China \\
\email{xywang@nju.edu.cn}
\and Guangxi Key Laboratory for Relativistic Astrophysics, Nanning 530004, China
\and Key laboratory of Modern Astronomy and Astrophysics (Nanjing University), Ministry of Education, Nanjing 210093, China
\and National Astronomical Observatories, Chinese Academy of Sciences, Beijing 100012, China\\
\email{dili@nao.cas.cn}
\and CAS Key Laboratory of FAST, NAOC, Chinese Academy of Sciences, Beijing,100101, China
}

\abstract {Star-forming regions on different scales, such as giant molecular clouds in our Galaxy and star-forming galaxies, emit GeV gamma-rays. These are thought to originate from  hadronic interactions of cosmic-ray (CR) nuclei with the interstellar medium.
It has recently been shown that the gamma-ray luminosity ($L_\gamma$) of star-forming galaxies is well correlated with their star formation rates (SFR). We investigated \textsl{Fermi} data of eight Galactic molecular clouds in the Gould belt and found that molecular clouds do not follow the $L_\gamma-{\rm SFR}$ correlation of star-forming galaxies. We also compared the scaling relations of gamma-ray luminosity, SFR, and the gas mass for  molecular clouds and star-forming galaxies.
Using a multiple-variable regression analysis,  we found different dependences of gamma-ray emission on SFR or mass for molecular clouds and star-forming galaxies. This suggests that different mechanisms may govern the production of gamma-rays in these two types of sources.
Specifically, the strong dependence on mass supports that gamma-ray emission of molecular clouds primarily comes from {\em \textup{passive}} interaction by diffuse Galactic CRs, whereas the strong dependence on SFR supports that  gamma-ray emission of star-forming galaxies originates from CRs that are accelerated by local {\em \textup{active}} sources.}
\maketitle

\section{Introduction}
Several nearby star-forming and starburst galaxies have been identified to be GeV to TeV gamma-ray
sources (e.g., \citet{2009Sci...326.1080A,2009Natur.462..770V,2010A&A...523L...2A,2012ApJ...755..164A,2014ApJ...794...26T,2016ApJ...821L..20P,2016ApJ...823L..17G,2017ApJ...836..208A}).
Cosmic rays (CRs) accelerated by supernova remnants (SNRs) or the stellar wind of massive stars interact with the interstellar medium (ISM) and produce neutron pions (schematically written as $p+p\rightarrow p+\pi^0+$ other products), which in turn decay into high-energy gamma-rays ($\pi^0\rightarrow\gamma+\gamma$).
Interestingly, with the early \textsl{Fermi} Large Area Telescope (LAT) data, \citet{2010A&A...523L...2A} found a correlation between the gamma-ray ($> 100 \rm \ MeV$) luminosity ($L_{\gamma}$) and star formation rate (SFR) for nearby star-forming galaxies.
Based on three years of \textsl{Fermi}-LAT data, a tight correlation between the gamma-ray luminosity and total infrared luminosity ($8-1000 \ \mu m$) over $4-5$ orders of magnitudes has been reported for star-forming galaxies by \citet{2012ApJ...755..164A}.
Since the total infrared luminosity is an indicator of the SFR of star-forming galaxies and galaxies that are not yet detected in gamma-rays are also taken into account to reduce sample selection effects, this then suggests that the positive correlation between the gamma-ray luminosity and SFR ($L_{\gamma}-\rm SFR $) is robust.
Lately, this correlation has been extended to ultra-luminous infrared galaxies, with the detection of gamma-ray emission from  Arp 220 \citep{2016ApJ...821L..20P,2016ApJ...823L..17G}. This strengthens the connection between star formation process and gamma-ray emission on a larger luminosity scale.

Molecular clouds, which are the sites of star formation in our Galaxy, are also sources of gamma-ray emission (e.g.,  \citet{2012ApJ...755...22A,2012ApJ...756....4A,2012ApJ...750....3A}). The widely accepted explanation of the gamma-ray emission of molecular clouds is that the clouds are passive targets for interaction with diffuse Galactic CRs (e.g., \cite{2001SSRv...99..187A,2005Sci...307.1292G,2007Ap&SS.309..365G}).
The passive-target scenario assumes that Galactic cosmic rays can freely penetrate the clouds and enter into the core region.
This scenario has gained support from the fairly uniform distribution of gamma-ray emissivity per gas nucleon in the Gould Belt clouds and in the Local Arm \citep{2009ApJ...703.1249A,2010ApJ...710..133A,2011ApJ...726...81A,2012ApJ...756....4A,2012ApJ...750....3A,2012AIPC.1505...37C}.

On the other hand, there have been suggestions that molecular clouds contain active sources of CRs. Embedded young stellar objects (YSOs) in giant molecular clouds are proposed to be sources of high-energy cosmic rays (e.g., \citet{2007A&A...476.1289A,2010A&A...511A...8B,2016A&A...591A..71M}).
The strong stellar wind activity in these objects generates large bubbles and induces collective effects that could accelerate particles up to high energy and produce gamma-rays (e.g., \citet{2011Sci...334.1103A,2017A&A...600A.107Y,2018A&A...611A..77Y,2018arXiv180402331A}).
For low-mass YSOs such as T Tauri protostars, their jets and winds should be unimportant. However, these objects have bubbles and outflows that seem to sustain turbulence \citep{2015ApJS..219...20L}. This would mean that particles could also be accelerated to relativistic energies by turbulence or a magnetic reconnection process \citep{2011ApJ...738..115D}.

It has been clear that stars are born in the molecular gas, especially in regions with dense molecular gas, rather than those primarily with atomic hydrogen. The dense region within molecular clouds collapses and then forms new stars. The SFR of molecular clouds estimated from the far-infrared emission correlates well with the dense molecular gas mass traced by $\rm HCN$ for galaxies (e.g., \citet{2004ApJ...606..271G}). This correlation continues to dense Galactic cores over a large scale of magnitude in SFR (e.g., \citet{2005ApJ...635L.173W,2010ApJ...724..687L}).
As molecular clouds are a scaled down version of star-forming galaxies in some sense, the question arises whether the $L_{\gamma} - {\rm SFR}$ correlation can extend to the scale of Galactic molecular clouds.

To study whether the SFR plays an important role in producing gamma-ray emission in molecular clouds, we here conduct a comparative study of the correlation between gamma-ray luminosity and SFR for Galactic molecular clouds and star-forming galaxies. Furthermore, we systematically investigate the correlations among $L_{\gamma}$, SFR, and gas mass $M$ to study the relative dependence of $L_{\gamma}$ on SFR and gas mass.
We analyzed the \textsl{Fermi}-LAT data of eight Galactic molecular clouds in the Gould Belt.
We noted that the \textsl{Fermi}-LAT data of local molecular clouds of the Gould Belt had been used to probe the CR properties by several groups \citep{2012PhRvL.108e1105N,2014A&A...566A.142Y,2017A&A...606A..22N,2018arXiv180106075S}. However, these works focused on deriving the spectrum of the parent CRs through the resolved gamma-ray spectral analysis and compared it with the measurements of local Galactic CRs.
Their relatively large offsets from the Galactic plane ensure little contamination from the diffuse Galactic gamma-ray emission.
The close distances lead to a high detection significance of the gamma-ray emission.
These make them good candidates for our study.
The structure of the paper is as follows. In Section 2 we describe the data reduction and results of the \textsl{Fermi}-LAT observations.
In Section 3 we study whether the $L_{\gamma} - {\rm SFR}$ correlation of galaxies can extend to the scale of Galactic molecular clouds. In Section 4 we present the study on the correlation among gamma-ray luminosity $L_{\gamma}$, SFR, and the gas mass $M$.
In Section 5 we discuss the implications of these two-parameter and three-parameter correlations.
Finally, we conclude in Section 6.

\section{\textsl{Fermi}-LAT data reduction}
The LAT on board the \textsl{Fermi} mission is a pair-conversion instrument that is sensitive to GeV emission \citep{2009ApJ...697.1071A}.
We accumulated data events from the start (MET 239557417) to 2017 April 6 (MET 513164606) to study gamma-ray emission from the Galactic molecular clouds. The basic information of the clouds, including the position, size, mass, and distance, is listed in Tables \ref{tableinfobasic} and \ref{tableinfofermi}.
The current \textsl{Fermi}-LAT Pass 8 SOURCE data (P8R2 Version 6) and the standard \textsl{Fermi} science tools version v10r0p5 were used.
All FRONT+BACK converting photons with energies higher than $0.3 \ \rm GeV$ were taken into consideration to reduce the contamination from poor angular resolution events at lower energies.
To limit the gamma-rays produced by CR interactions in the upper atmosphere, the maximum zenith-angle cut $zmax = 90^{\circ}$ was required.
The expression of (DATA\_QUAL $> 0$) \& \& (LAT\_CONFIG ==1) was used to further filter the data in the \textsl{gtmktime}.
We performed the binned maximum likelihood analysis on a region of interest (ROI) with a radius of $10^{\circ}$ centered on the position of each Galactic molecular cloud.
We used the 3FGL \citep{2015ApJS..218...23A} to generate the source model containing the position and spectral definition for all the point sources and diffuse emission within $15^{\circ}$ of the ROI center. The extragalactic diffuse model \textsl{iso\_P8R2\_SOURCE\_V6\_v06.txt} was also included.

All molecular clouds in our sample extend several degrees above the Galactic plane on the sky.
A comparison of the gamma-ray counts map with CO intensity reveals a good correlation between the gamma-ray and CO emission.
The CO distribution in the direction of sight was reduced to a single peak by integrating the CO cube in the spatial dimension \citep{2001ApJ...547..792D}.
We particularly focused on the integral gamma-ray flux of each molecular cloud for the correlation studies.
Because the Galactic diffuse gamma-ray emission in the ROI is dominated by the interaction between CRs and molecular clouds and neutral hydrogen is distributed diffusively in our Galaxy,  we adopted the following method. The templates for modeling the spatial and spectral distribution of molecular clouds were extracted from the standard Galactic diffuse emission model, that is, a cube file named as \textsl{gll\_iem\_v06.fits}, provided by the \textsl{Fermi}-LAT Collaboration.
We selected one region nearby with the same size for each molecular cloud as the background without significant CO emission. In each energy band, we obtained the average value in the background region, which was also used to model the residual Galactic diffuse emission of the molecular cloud. Then we obtained the molecular cloud templates by subtracting the above value from the standard \textsl{gll\_iem\_v06.fits}. After the source model was created, the standard commands {\em gtbin}, {\em gtltcube}, {\em gtexpcube2}, and {\em gtsrcmaps} were successively executed, then a maximum likelihood analysis was performed in binned mode using the tool {\em gtlike} \footnote{\url{https://fermi.gsfc.nasa.gov/ssc/data/analysis/scitools/binned_likelihood_tutorial.html}}.

Cosmic rays  generate  diffuse gamma-ray emission by interacting with interstellar gas and magnetic fields during their propagation through the Galaxy. One way to derive the spatial and spectral information of the diffuse gamma-ray emission templates is to use the GALPROP code \footnote{\url{https://galprop.stanford.edu/}}.
We therefore checked the results on a diffuse emission model that did not include a gas component with the help of GALPROP.
We used models for the predicted Galactic diffuse gamma-ray emission obtained from the \textsl{Fermi}-LAT collaboration work \citep{2012ApJ...750....3A}, which provides 128 sets of maps corresponding to different model parameters. We tested 16 of these 128 templates and chose two each for the CR source distribution (Lorimer pulsars\citep{2006MNRAS.372..777L}, SNRs\citep{1998ApJ...504..761C}), vertical boundaries (4 kpc and 10 kpc), spin temperature for the optical depth correction (150 K and $10^5$ K), and E(B-V) magnitude cut (2 mag and 5 mag). We did not include the $\rm H_2$ component in these Galprop simulations. We used the spatial templates for pion-decay, bremsstrahlung radiation, and inverse Compton gamma-rays generated by GALPROP \footnote{\url{https://galprop.stanford.edu/webrun.php}} \citep{2011CoPhC.182.1156V} to replace the background model. The gamma-ray emissions of molecular clouds using the above 16 background models are consistent with those in Table \ref{tableinfofermi}, and the fluxes change at most by $\sim 15\%$ in these models; see Table \ref{tablerates}. These small differences do not affect our statistical results and conclusion.

The integral gamma-ray fluxes of molecular clouds are presented in Table \ref{tableinfofermi}.
To conform with the energy range of the data for the star-forming galaxies, we extrapolated the flux in $0.3-100\ \rm GeV$ to that in the energy range of $0.1-100\ \rm GeV$. The statistical error is small due to the high-significance detection. The data of gamma-ray emission in $ 0.1-100 \ \rm GeV$ from star-forming galaxies were taken from previous publications \citep{2012ApJ...755..164A,2014ApJ...794...26T,2016ApJ...821L..20P}.

In order to reduce the possible contribution from leptonic emission, such as inverse Compton scatter and bremsstrahlung radiation at low energies, which would overestimate the flux due to hadronic CRs interaction, we also considered the \textsl{Fermi}-LAT data in $ 1-500 \ \rm GeV$. Moreover, the possible gamma-ray contamination emission from unresolved sources such as pulsars was also suppressed by increasing the threshold energy for the data analysis \citep{2013ApJS..208...17A}. To obtain the gamma-ray luminosity in 1-500 GeV ($L_{1-500\ \rm GeV}$), we performed a likelihood analysis of the latest \textsl{Fermi}-LAT data following a method that was similar to the previously used method.
The results are shown in Tables \ref{tableinfofermi} and \ref{tableinfoGax}.

\section{Do molecular clouds follow the $L_{\gamma}$-SFR correlation of star-forming galaxies?  }
As molecular clouds are a scaled-down version of star-forming galaxies in some sense, we first studied whether the $L_{\gamma} - {\rm SFR}$ correlation of galaxies can extend to the scale of Galactic molecular clouds. To do this, we checked whether the Galactic molecular clouds fall onto the correlation line in the  $L_{\gamma} - \rm SFR$ diagram of star-forming galaxies.
Since $L_{\gamma}$ and SFR of molecular clouds and star-forming galaxies span $7-8$ orders of magnitude, we reduced the dynamic range from clouds to star-forming galaxies by dividing the gamma-ray luminosity and SFR  by the gas mass $M$. We compared the gamma-ray emissivity ($L_\gamma/M$) and SFR per unit mass for molecular clouds and star-forming galaxies.
Although the methods for determining the gas masses for molecular clouds and star-forming galaxies are different, it has been proved that the molecular-line-derived masses and the extinction-derived masses accurately reflect the same material \citep{2012ApJ...745..190L}. We computed the average integral $> 100 \ \rm MeV $ gamma-ray emissivity per hydrogen atom of molecular clouds using the following form: $q_{\gamma} = 8.0\times 10^{-27} \frac{F_{\gamma}}{10^{-7} \ \rm ph \ cm^{-2} \ s^{-1}}(\frac{d}{1 \ \rm kpc})^2 (\frac{M_{\rm gas}}{10^5 \ M_{\odot}})^{-1}$, where the emissivity $q_{\gamma}$ is in unit of $\rm ph\ s^{-1}\ sr^{-1}\ H^{-1}$, $F_{\gamma}$ is the integral photon flux, $M_{\rm gas}$ is the total gas content of the molecular cloud, and $d$ is the distance to Earth. The mean value of the ratio between the measured integral gamma-ray emissivity of local atomic hydrogen \citep{2009ApJ...703.1249A} and that of the sample molecular clouds in our work is $1.13\pm 0.69$, and the median value is $0.90$, indicating that the gamma-ray emissivities of molecular clouds are quite close to the emissivity of local atom hydrogen.

\begin{figure*}
\centering
\includegraphics[scale=0.4]{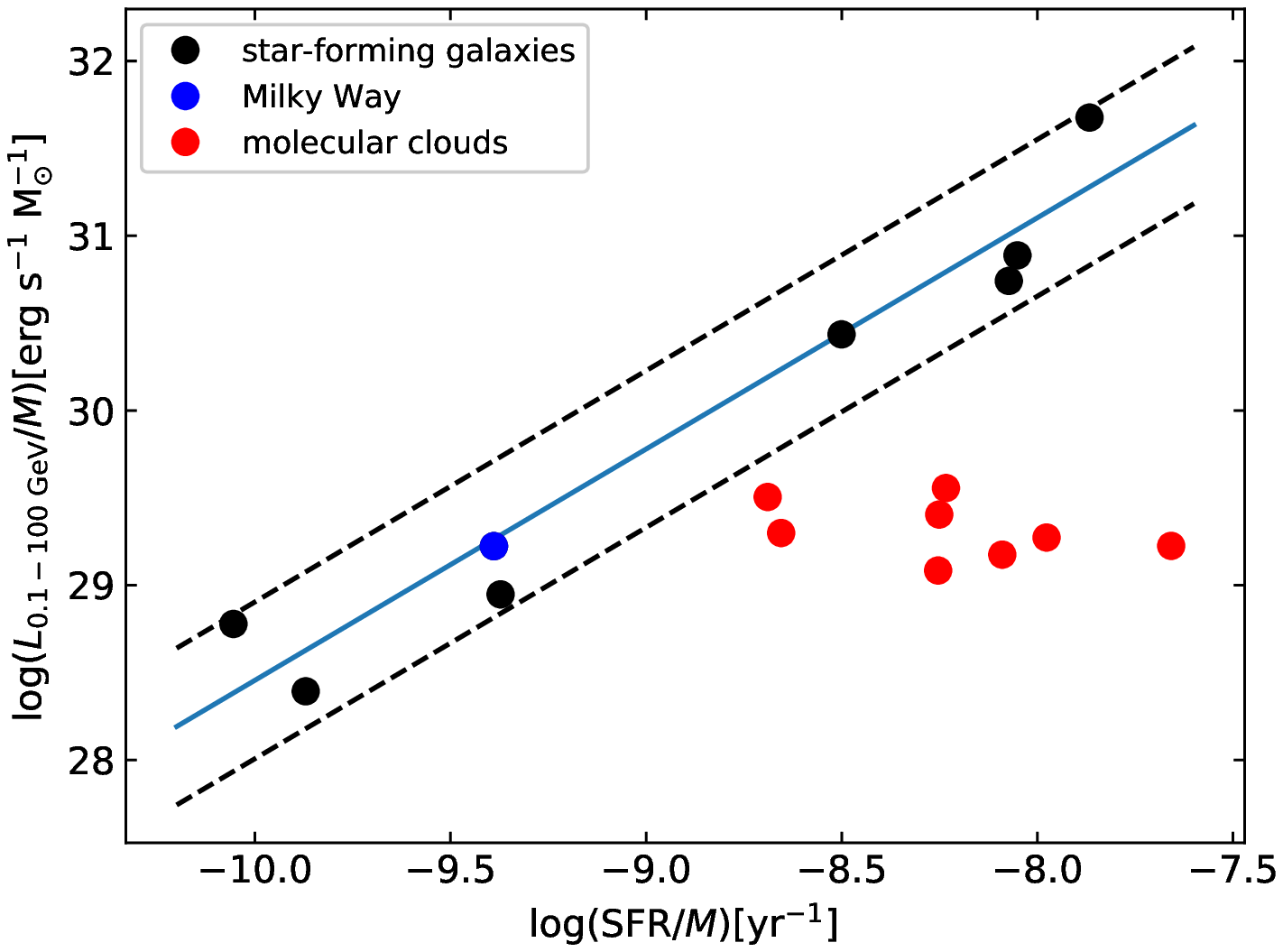}
\includegraphics[scale=0.4]{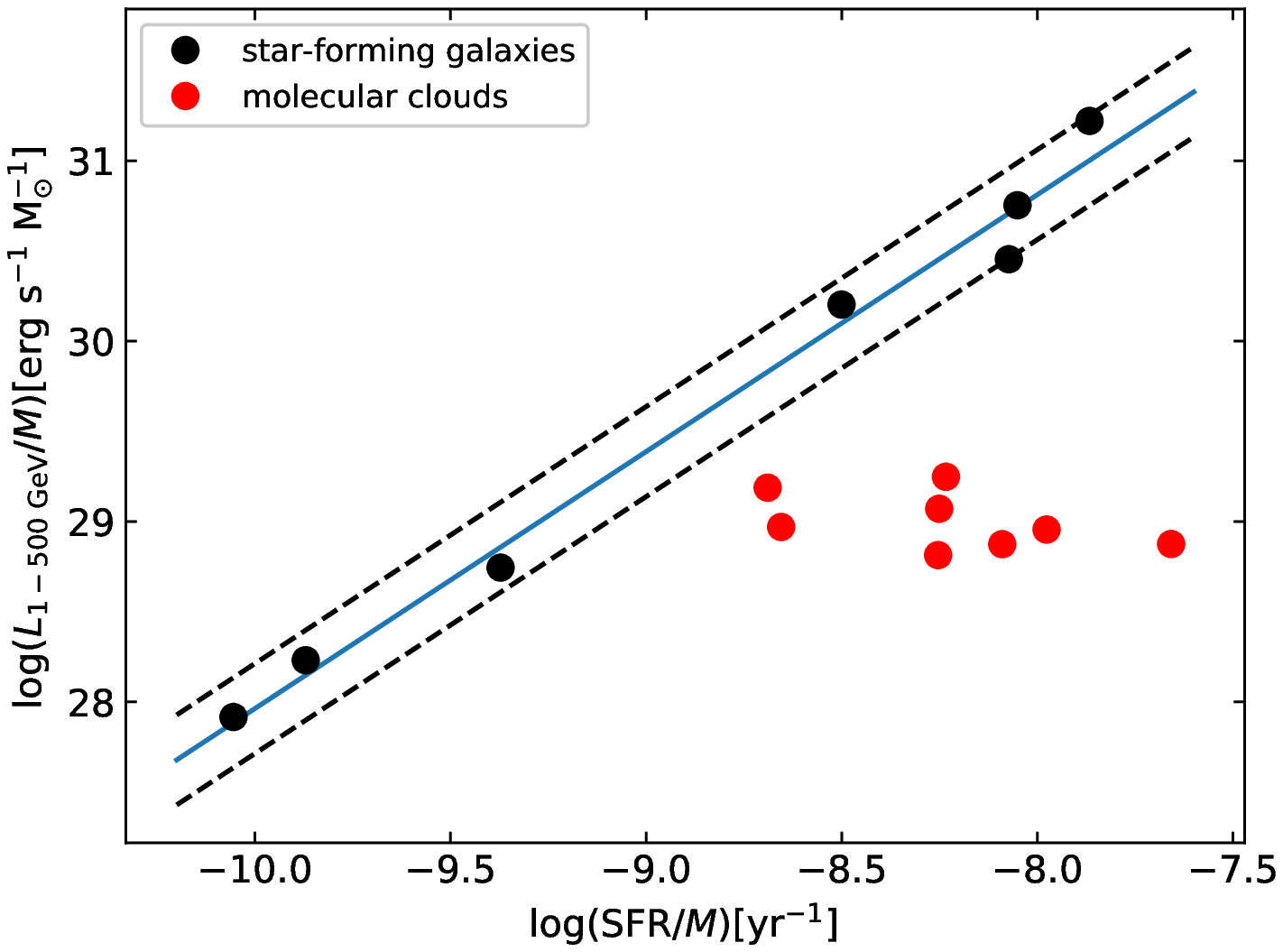}
\caption{Relation between gamma-ray emissivity (left panel: $0.1-100 \ \rm GeV$; right panel: $1-500 \ \rm GeV$) and SFR per unit mass for molecular clouds and star-forming galaxies. The best-fit lines together with their $1\sigma$  dispersion regions are shown with solid and dashed lines, respectively.}
\label{figratio}
\end{figure*}

The results are reported in Figure \ref{figratio}. A correlation is evident between the gamma-ray emissivity ($L_\gamma/M$) and SFR per unit mass for star-forming galaxies,  but the molecular clouds  significantly deviate from this correlation.
The Pearson correlation coefficient of star-forming galaxies is $r > 0.9$ and the chance probability is $p < 10^{-4}$. The derived total dispersions, including the intrinsic scatter of the data set and the statistic scatter, are 0.45 and 0.24 for the gamma-ray emissivity ($L_\gamma/M$) in 0.1-100  GeV and 1-500 GeV, respectively.
The gamma-ray emissivities ($L_\gamma/M$) of molecular clouds are distributed in a narrow space and are comparable to the mean value of the Milky Way, which contains a great variety of different molecular clouds with different star-forming activities.
The different scalings imply that the gamma-ray emission of molecular clouds and star-forming galaxies has a different origin. The roughly constant gamma-ray emissivity among the clouds as well as the Milky Way supports the hypothesis that the gamma-ray emission of clouds is due to passive  interactions by the diffuse Galactic CRs.
This wide range of $\rm SFR/M$ may reflect the variations in the fractions of dense gas. The physical interpretation behind it could be due to the cloud evolution. The evolution of molecular clouds is controlled by a complex interplay of large-scale phenomena and microphysics, such as turbulence, magnetic field, outflow of young stellar objects, far-ultraviolet radiation, CR radiation, gas, and dust (e.g., \citet{2011MNRAS.414.2511V}, and reference therein).

\section{Multiple-variable regression analysis}
In this section we conduct a multiple-variable regression analysis to study the correlations among the parameters of gamma-ray luminosity, SFR, and $M$ for molecular clouds and star-forming galaxies, respectively. The underlying theory for this statistical analysis is that gamma-ray emission should depend on both the sources of CRs and
the target gases. The sources of Galactic CRs could be SNRs and/or young stellar objects, which are indicated by the SFR. We therefore studied the correlation between gamma-ray luminosity and SFRs for molecular clouds and star-forming galaxies. Furthermore, we systematically
investigated the correlations among  gamma-ray luminosity, SFR, and gas mass M to study the relative dependence of gamma-ray emission on
SFR and gas mass. Through a comparative and statistic analysis, we examine whether the SFR plays an import role in producing gamma-ray emission
in molecular clouds, and we study the difference of gamma-ray emission processes in molecular clouds and star-forming galaxies.

Generally speaking, the results of the regression analysis depend on the choice of dependent and independent variables \citep{1990ApJ...364..104I,1992ApJ...397...55F}, especially in our case, where the sample size is small and the intrinsic data scatter is large.
The bisector or orthogonal method could be adopted to solve the problem from the point of view of mathematics. However, from a physical point of view, the independent and dependent variables are believed to be clear.
For the given data set, SFR stands for the sources of CRs, the gas mass $M$ stands for the target material, and $L_{\gamma}$ is the result of CR interaction between sources \textup{}and target\textit{}. Therefore, $L_{\gamma}$ was chosen as the independent variable in the following analysis.

The two-parameter correlation between $L_{\gamma}$ and SFR for nearby star-forming galaxies has been found for the first time by \citet{2010A&A...523L...2A}, and it has been confirmed by follow-up studies \citep{2012ApJ...755..164A,2014ApJ...794...26T,2016ApJ...821L..20P,2016ApJ...823L..17G}.
This discovery is linked to the relation between CRs and SFR, although its origin is not yet fully understood.
Here we explore the two-parameter correlations between gamma-ray luminosity and SFR (or gas mass) for star-forming galaxies and clouds to determine the roles that the SFR plays in producing gamma-ray emission and to understand the physical nature behind the scaling relations.

We modeled the two-parameter correlation using the form $z = a + b\times x$.
To obtain the best-fitting parameters to the observational data with the two-parameter correlation analysis, we used the maximum likelihood approach.
The joint likelihood function for two-parameter analysis is
\begin{equation}
\mathcal {L}(a,b,\sigma) = \prod\limits_{i} \frac{1}{\sqrt{2 \pi \sigma^2}}\times e^{-\frac{(z_{i}- a -b\times x_{i} )^2}{2\sigma^2}},
\end{equation}
where $i$ is the corresponding serial number of molecular clouds or star-forming galaxies in our sample, $z$ is $L_{\gamma}$, and $x$ is SFR (or $M$). They all are derived in logarithmic space.
Since no error bars for SFR and $M$ are available, and very small statistic errors in $L_{\gamma}$  for some molecular clouds (see Table \ref{tableinfofermi}) would lead to relatively large weights,  no measurement errors were considered.
$\sigma$ is introduced to accommodate intrinsic scatter and measurement errors.
The coefficients of $a, \ b,$ and $\sigma$ are constrained simultaneously by maximizing the joint likelihood function.

For each sample of molecular clouds and star-forming galaxies, we used the Python Markov chain Monte Carlo module
EMCEE \citep{2013PASP..125..306F} to explore the posterior distributions of parameters of $a, \ b,$ and $\sigma$. We derived the dispersion ($\delta$) of a regression model with standard deviation of $z^{\rm r}$ from $z$, where $r$ marks the $z$ value derived from the regression model.

\begin{figure*}
\centering
\includegraphics[scale=0.4]{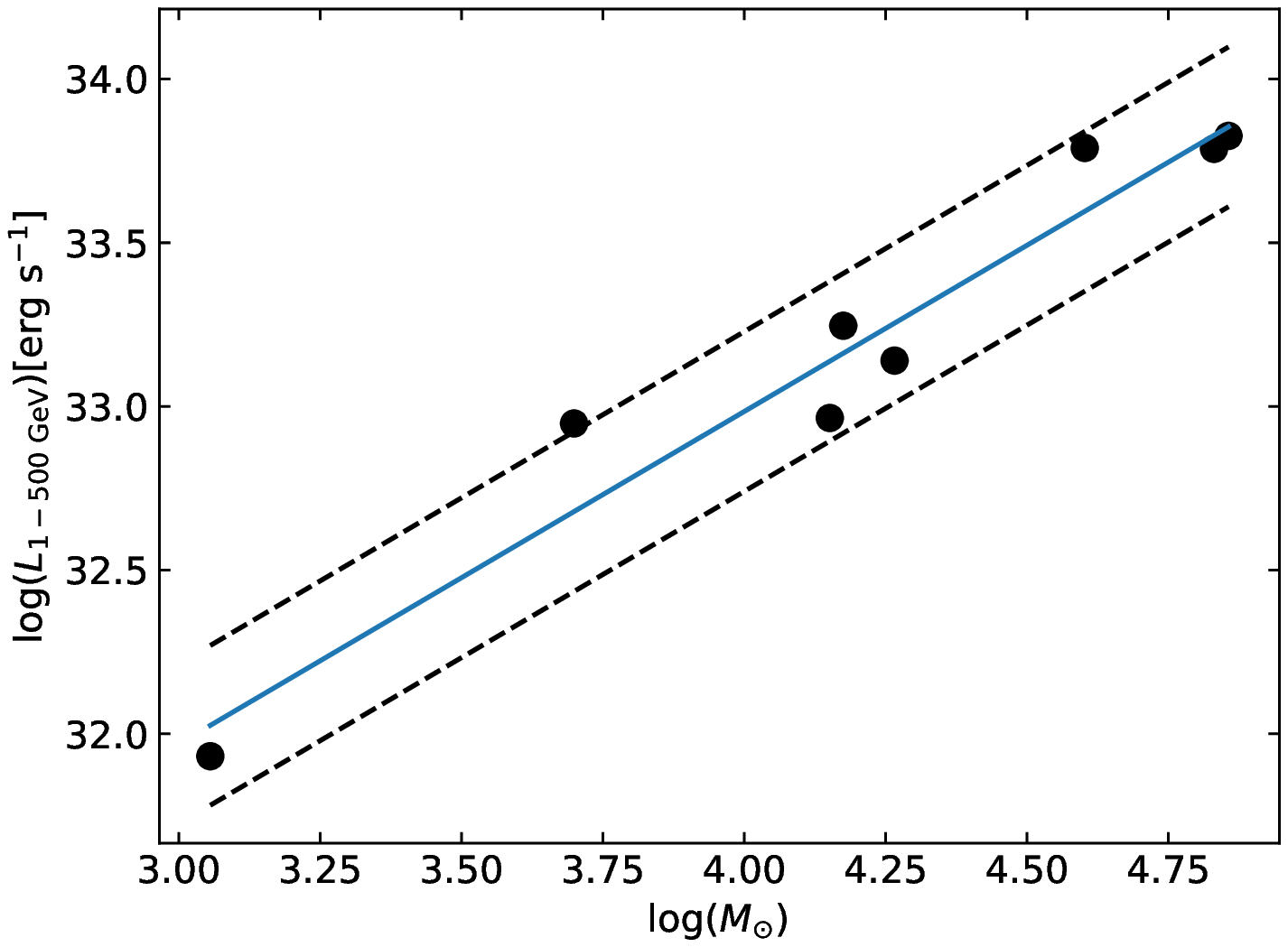}
\includegraphics[scale=0.4]{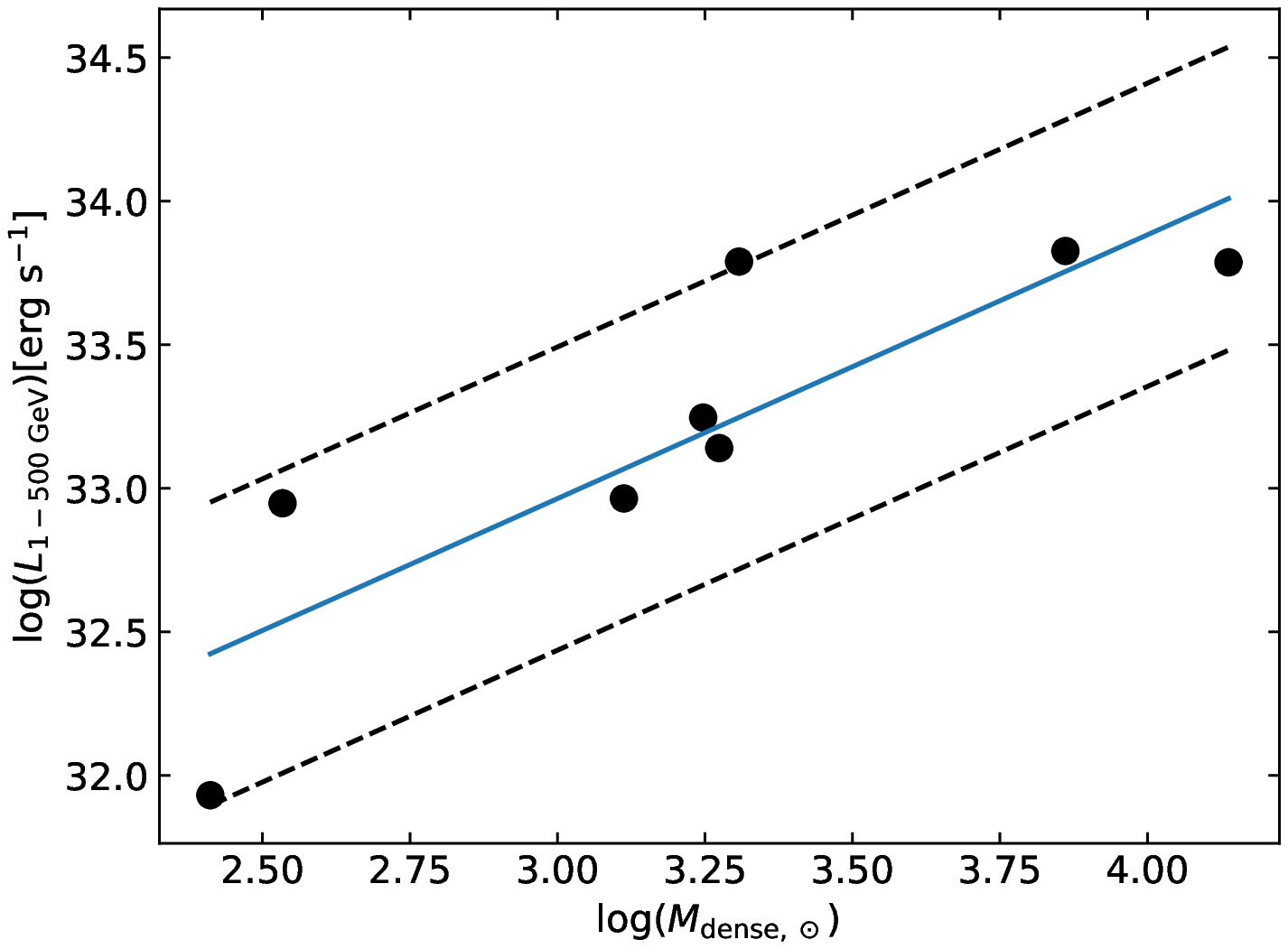}
\caption{Two-parameter correlation between $L_{1-500 \ \rm GeV}$ and total gas mass (left panel) and dense gas mass (right panel) for molecular clouds. The best-fit lines together with their $1\sigma$  dispersion regions are shown with solid and dashed lines, respectively.}
\label{figMCgamMass}
\end{figure*}

We applied this two-parameter correlation analysis to  molecular clouds and star-forming galaxies using the forms ${\rm log}L_{\gamma} = a + b{\rm log(SFR)}$ or ${\rm log }L_{\gamma} = a + b{\rm log}(M)$. The results of the correlations are shown in Tables \ref{tableLumSFR} and \ref{tableLumM}.
The results in Tables \ref{tableLumSFR} and \ref{tableLumM} show at first glance that all correlations between $L_{1-500 \ \rm GeV}$ and SFR or $M$  are significant from the statistical point of view, even though there is some difference in the correlation coefficients and the dispersion of the fit.
Particularly, stronger dependences of $L_{1-500 \ \rm GeV}$ on gas mass $M$ for molecular clouds and $L_{1-500 \ \rm GeV}$ on SFR for star-forming galaxies are indicated by the Pearson correlation coefficient $r > 0.9$ and the chance probability $p < 10^{-4}$. Figure \ref{figMCgamMass} shows the correlation between gamma-ray luminosity and mass $M$ of the molecular cloud sample.

As described above, the gamma-ray emission from star-forming regions originates from CR interactions. In principle, $L_{\gamma}$ may depend  not only on the number of CR sources denoted by the SFR, but also on the mass of the target gas that is denoted by $M$. Therefore, it is interesting to investigate the possible multi-parameter correlation among $L_{\gamma}$, SFR, and $M$. The likelihood function can also be conveniently applied to the three-parameter correlation case by introducing an additional term of $c\times y_i$, that is, ${\rm log}L_{\gamma} = a + b{\rm log(SFR)} + c{\rm log}(M) $.
The relative dependence of $L_{\gamma}$ on SFR or $M$  is shown clearly through this three-parameter correlation analysis, which is helpful to reveal the mechanism of gamma-ray emission.

\begin{figure*}
\centering
\includegraphics[scale=0.40]{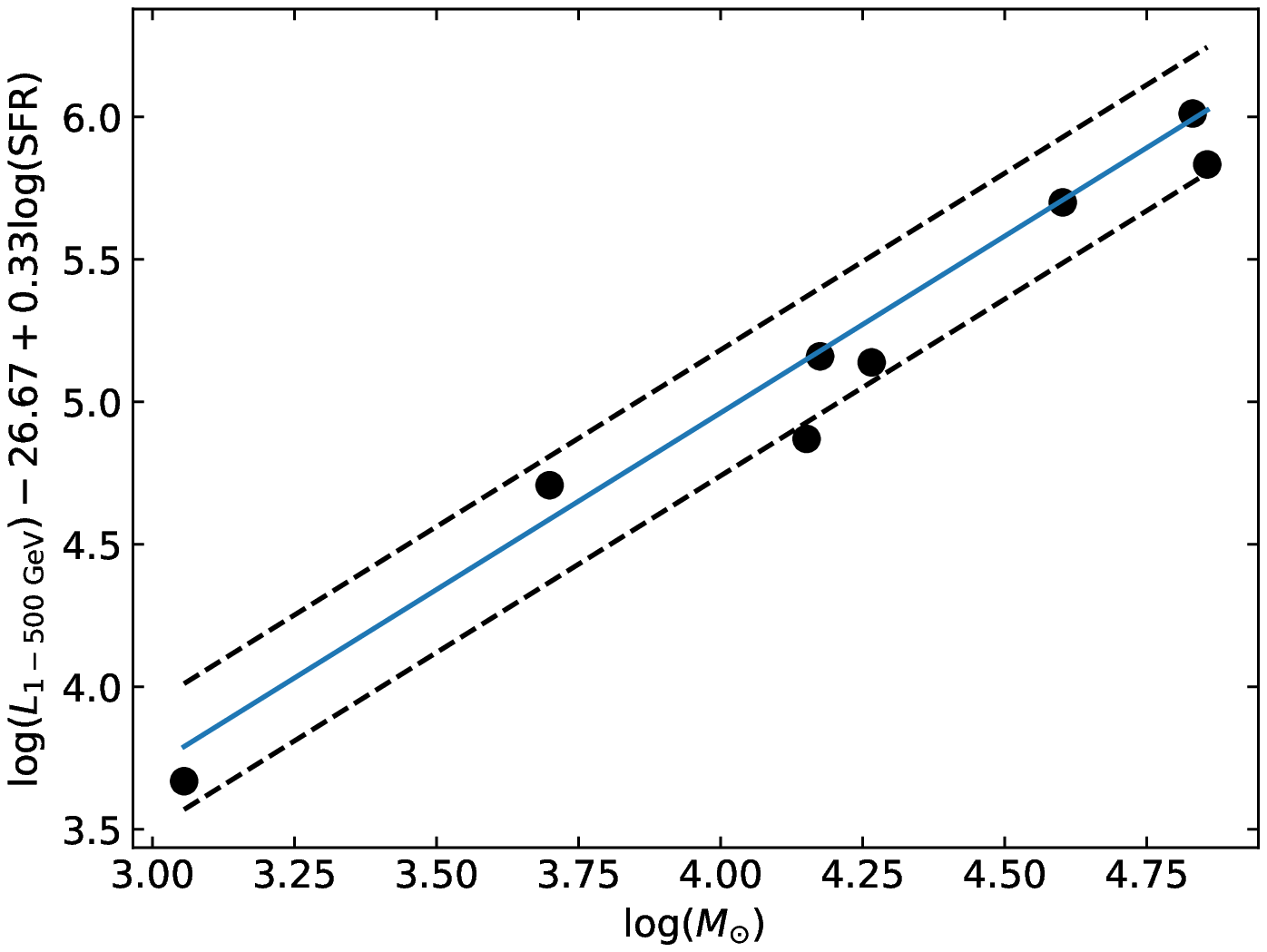}
\includegraphics[scale=0.40]{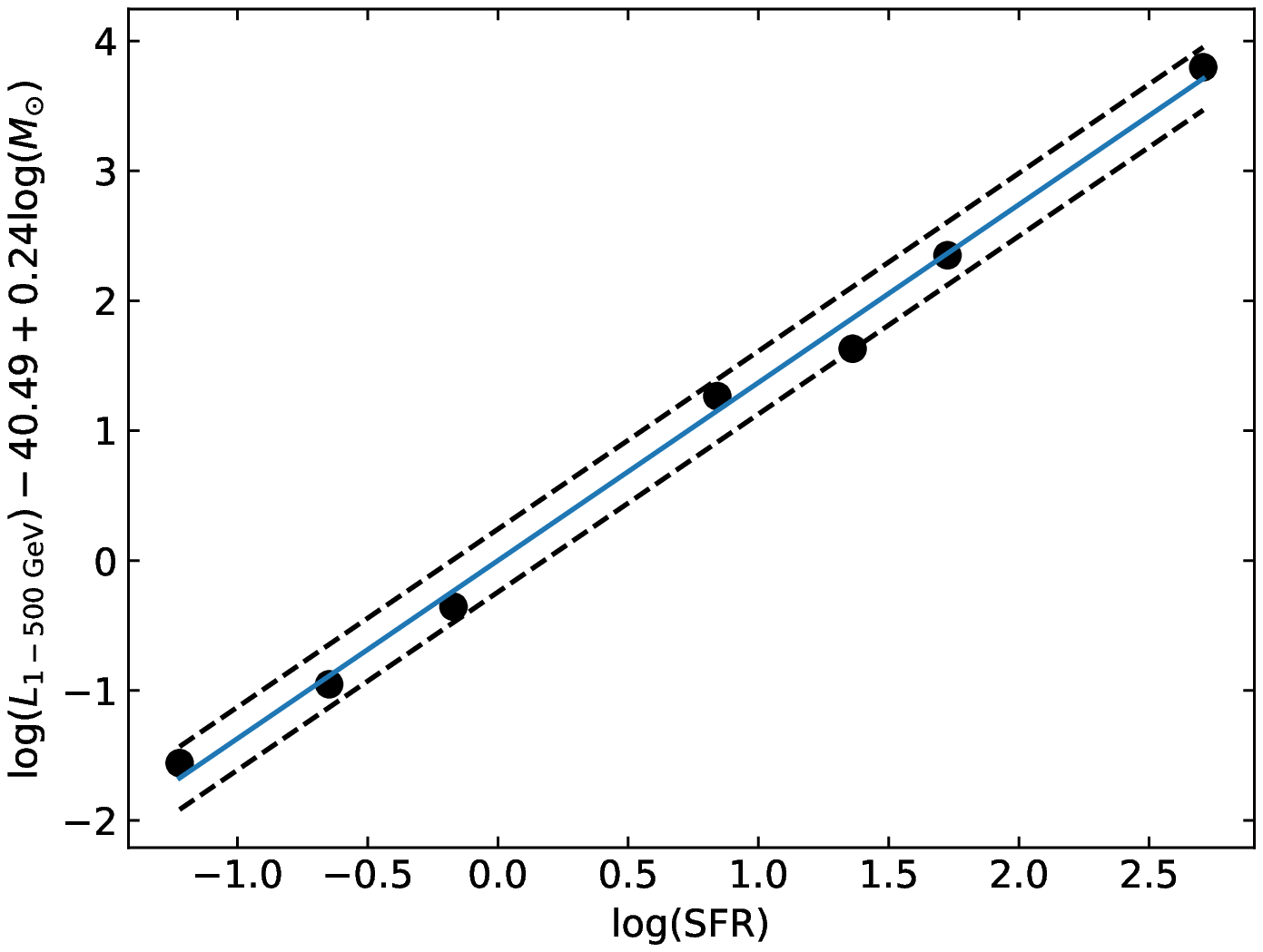}
\caption{Three-parameter correlation among $L_{1-500 \ \rm GeV}$, SFR, and $M$ for molecular clouds (left panel) and star-forming galaxies (right panel). The best-fit lines together with their $1\sigma$  dispersion regions are shown with solid and dashed lines, respectively.}
\label{figthree}
\end{figure*}

The results of three-parameter correlations among $L_{\gamma}$, SFR, and $M$ for  molecular clouds and star-forming galaxies are reported in Table \ref{tableGeV}. The best-fit correlations are
\begin{equation}\label{MCs}
\begin{split}
{\rm log}(L_{1-500 \ \rm GeV}) = (26.67^{+2.02}_{-1.90})+ (-0.33^{+0.29}_{-0.29}){\rm log(SFR)} \\
+ (1.23^{+0.22}_{-0.22}){\rm log}(M)
\end{split}
\end{equation}
for molecular clouds and
\begin{equation}\label{Gax}
\begin{split}
{\rm log}(L_{1-500 \ \rm GeV}) = (40.49^{+2.24}_{-2.25})+(1.37^{+0.12}_{-0.12}){\rm log(SFR)}\\
 + (-0.24^{+0.25}_{-0.25}){\rm log}(M)
\end{split}
\end{equation}
for star-forming galaxies, respectively.  The derived $1\sigma$ errors of these coefficients are listed in Table \ref{tableGeV}.
The Pearson correlation coefficient $r > 0.9$ and chance probability $p < 10^{-4}$ suggest  strong correlations among $L_{1-500 \ \rm GeV}$ , SFR, and $M$.
For molecular clouds, the weak dependence on SFR, as indicated by $b \sim 0 $ within the error box, shows that the $1-500\ \rm GeV$ gamma-ray luminosity is principally proportional to the gas mass $M$.
However, the results for galaxies are the opposite. $c = -0.24^{+0.25}_{-0.25}$ means that the gamma-ray luminosity of galaxies depends little on the total gas mass $M$. There is a clear trend that $L_{1-500 \ \rm GeV}$ increases with SFR.
These results are consistent with the two-parameter correlations analysis.
The dispersion and linear coefficient  for the three-parameter correlation are almost the same as those of the two-parameter fit  of $L_{1-500 \ \rm GeV} -  M $ for molecular clouds and $L_{1-500 \ \rm GeV} - \rm SFR $ for star-forming galaxies, respectively.
Introducing the third parameter into the three-parameter correlation does not improves the fit  significantly, which implies a weak dependence of $L_{\gamma}$ on SFR for molecular clouds, and  on $M$ for star-forming galaxies. The regression lines together with $1\sigma$ dispersion regions are also presented in Figure \ref{figthree}.
For a comparison with previous works, we also performed a multi-parameter correlation analysis for molecular clouds and star-forming galaxies using a  $0.1-100$ GeV gamma-ray luminosity. The results are reported in Table \ref{tableMeV}

\section{Discussion }
The correlation among $L_{\gamma}$ , SFR, and $M$ is of theoretical interest in understanding the mechanism of GeV emission in molecular clouds and star-forming galaxies.
The formula for estimating the gamma-ray emission quantitatively \citep{2018arXiv180402331A} reads
\begin{equation}
\frac{L_{\gamma}(\geqslant  {E_{\gamma}})}{10^{34} \ \rm {erg \ s^{-1}}} = 5.6 (\frac{M}{10^{5} M_{\odot}})(\frac{\eta}{1.5})(\frac{u_{\rm CR}(\geqslant 10 E_{\gamma})}{1 \ \rm{ev \ cm^{-3}}}),
\end{equation}
where $M$ is the mass of the relevant region, $\eta$ accounts for the presence of nuclei higher in mass than
hydrogen in CRs and interstellar matter, and $u_{\rm CR}$ is the CR density.

If the molecular clouds in our sample are an active source of CR acceleration, similar to star-forming galaxies, or if the embedded YSOs can contribute CRs at a comparable level of the Galactic CR sea, gamma-ray emission should also show a correlation with CR density (denoted as SFR), smoothly connecting the $L_{\gamma} -\rm SFR$ relationship of star-forming galaxies.
Correspondingly, the dependence of $L_{\gamma }$ on the SFR in three-parameters correlation should be much stronger. However, this is not supported by our data analysis. The data set of molecular clouds and star-forming galaxies is located in  different regions  in the diagram of gamma-ray emissivity ($L_\gamma/M$) and SFR per unit mass (Figure \ref{figratio}).
The gamma-ray emissivity ($L_\gamma/M$) remains to be a constant for different ${\rm SFR}/M$ for Galactic molecular clouds. This is consistent with the gamma-ray's being produced outside of the cloud and with the recent study of diffuse \textsl{Fermi} gamma-ray emission, which seems to trace the total molecular gas content on a global scale \citep{2018A&A...611A..51R}.
The nice linear correlation between $L_{\gamma}$ and total mass for molecular clouds is demonstrated by the slope of $1.02^{+0.14}_{-0.14}$.
In other words, the CRs that produce the dominant part of the gamma-ray emission in molecular clouds may be accelerated outside, as expected from the passive-cloud scenario.
These molecular clouds float in the sea of the Galactic CRs, and the produced gamma-ray emission is proportional to the total gas mass under the hypothesis that the non-violent change CRs flux penetrates the clouds.
The embedded massive stars may contribute to CRs secondarily or account for some CR \textsl{\textup{hot spots}} around clusters of young stellar objects (e.g., \citet{2018arXiv180305968M}).
The Pearson correlation coefficient of $L_{\gamma} -\rm SFR$ for molecular clouds is also good from a statistical point of view, but it may simply reflect the $L_{\gamma} -M_{\rm dense} $ correlation, as $M_{\rm dense}$ is a proxy of SFR.
It could be naturally explained by the combination of an only mildly varying fraction of dense gas mass (see Table \ref{tableinfobasic}) and a tight $L_{\gamma} - M$ correlation.

For star-forming galaxies, the Pearson correlation coefficient and dispersion of $L_{\gamma} - M$ correlation demonstrates that the two-parameter relationship is poor, which is consistent with results of the three-parameter correlation.
The gamma-ray luminosity of star-forming galaxies can be parameterized by $L_\gamma\sim f_{\rm cal} L_{\rm CR}$, where $L_{\rm CR}$ is the CR luminosity in galaxies, which is proportional to the SFR, and $f_{\rm cal}$ is the  calorimetric factor denoting the fraction of the energy of CRs converted into secondary pions.
Star-forming galaxies with an SFR $>10{\ M_\odot \rm \ yr^{-1}}$ are close to the calorimetric limit (e.g., \citep{2011ApJ...734..107L,2018MNRAS.474.4073W}), that is,  $f_{\rm cal}\simeq 1$.
If the CR calorimetry hypothesis were to hold, the slope of the relation between the gamma-ray luminosity and SFR would be unity.
The observed slope of the relation is steeper than unity, however \citep{2010A&A...523L...2A,2012ApJ...755..164A,2014ApJ...794...26T,2016ApJ...821L..20P}, indicating that  galaxies with lower SFRs may have  smaller $f_{\rm cal}$. \citet{2017ApJ...847L..13P} reproduced the observed relation between far-infrared and gamma-ray emission using  magnetohydrodynamical galaxy formation simulations with self-consistent CR physics.
They found that the calorimetric factor $f_{\rm cal}$  decreases smoothly toward lower SFRs due to the increasing adiabatic losses of CRs.

\section{Conclusions}
We have analyzed the \textsl{Fermi}-LAT data of eight Galactic molecular clouds in the Gould Belt. Through a comparative study of the correlations among the gamma-ray luminosity, SFR, and gas mass $M$, we found that the gamma-ray luminosity of  molecular clouds is strongly dependent on the total gas mass $M$ and weakly dependent on SFR. The SFR inside molecular clouds makes minor contribution to the gamma-ray emission.
The results of star-forming galaxies are just the opposite. A tight dependence between the gamma-ray luminosity and SFR is found, with little dependence on $M$.
The different empirical correlations found in molecular clouds and star-forming galaxies indicates that different mechanisms produce the gamma-ray emission, with Galactic clouds being more of a passive target to interact with CRs.
The gamma-ray emission in molecular clouds originates predominantly from the interaction of diffuse galactic CRs. Star-forming galaxies are effective reservoirs for CRs, and a significant fraction of CR energy is transferred into secondary gamma-rays, at least for those with  GeV emission that are observed by \textsl{Fermi}-LAT.

\section*{Acknowledgments}

This work is partially supported by the National Key R \& D program of China under the grant 2018YFA0404203 and 2017YFA0402600, and the NSFC grants 11625312, 11851304, U1731238, 11565010, and 11725313. F.K. Peng acknowledges support from the Doctoral Starting up Foundation of Guizhou Normal University 2017 (GZNUD[2017] 33) and the open project of Guangxi Key Laboratory for Relativistic Astrophysics. Q.J. Zhi acknowledges support from the science and technology innovation talent team (grant (2015)0415), the High Level Creative Talents (grant (2016)-4008) and Innovation Team Foundation of the Education Department (grant [2014]35) of Guizhou Province. D. Li acknowledges support from the International
Partnership Program of Chinese Academy of Sciences, Grant No.114A11KYSB20160008 and the CAS Strategic Priority Re search Program No. XDB23000000.
We thank R.Z. Yang, R.-Y. Liu, J.Z. Wang, K.P. Qiu and H. Chen for useful discussions. This work has made use of data and software provided by the Fermi Science Support Center.

\clearpage

\begin{table*}
\centering
\caption{Basic information on Galactic molecular clouds. }
\begin{tabular}{lcccc}
\hline
\hline
Name  & Distance& $ M_{\rm total, \odot}$  & $ M_{\rm dense, \odot}$  & SFR \\
  & [pc]&   &   &  [$10^{-6}  \ M_{\odot} \ \rm yr^{-1}$] \\
\hline
RCrA    & $148\pm 30$\tablefootmark{1} & 1137 & 258 & 25\\
Oph     &       $119\pm 6$\tablefootmark{2}  & 14165&1296 & 79\\
Perseus &       $240\pm 13$\tablefootmark{3} & 18438&1880 &150
\\
Taurus  &       $153\pm 8$\tablefootmark{3} & 14964&1766 & 84\\
Orion A &       $371\pm 10$\tablefootmark{4}  & 67714& 13721 & 715\\
Orion B &       $398\pm 12$\tablefootmark{4} & 71828&7261 &159 \\
Chamaeleon      &       $200$\tablefootmark{5} & 5000\tablefootmark{5} & 342\tablefootmark{6} &29\tablefootmark{7}\\
Mon R2  &       $830$\tablefootmark{8} & 40000\tablefootmark{8} & 2031\tablefootmark{8}&82\tablefootmark{9}\\
\hline
\end{tabular}
\tablefoot{The third and forth columns are the total masses and dense masses estimated from the infrared extinction map at $A_K \geq 0.1$ mag and $A_K \geq 0.8$ mag, respectively \citep{2010ApJ...724..687L}. The last column is the SFR derived from the YSO observation. The masses and SFRs not marked are from \citet{2010ApJ...724..687L}.}\\
\textbf{References.} \tablefoottext{1}{\citet{2010arXiv1006.3676K}}; \tablefoottext{2}{\citet{2008A&A...480..785L}}; \tablefoottext{3}{\citet{2010A&A...512A..67L}}; \tablefoottext{4}{\citet{2011A&A...535A..16L}}; \tablefoottext{5}{\citet{2008hsf2.book..169L}}; \tablefoottext{6}{\citet{1999PASJ...51..859M}}; \tablefoottext{7}{\citet{2010ApJ...723.1019H}}; \tablefoottext{8}{\citet{2008hsf1.book..899C}}; \tablefoottext{9}{\citet{2016MNRAS.461...22P}}.

\label{tableinfobasic}
\end{table*}

\begin{table*}
\centering
\caption{Parameters and gamma-ray fluxes of Galactic molecular clouds.}
\begin{tabular}{lccccccc}
\hline
\hline
Name & ($l_s[^{\circ}]$,$b_s[^{\circ}]$)    & $\theta[^{\circ}]$ & ($l_b[^{\circ}]$,$b_b[^{\circ}]$) & $\rm Flux_{0.1-100 \ GeV}$ & Error & $\rm Flux_{1-500\ GeV}$ & Error \\
\hline
RCrA    &       (0.56,-19.63)   &       3       &       (6.94,-19.63)   &  0.91& 2.24  &  0.87& 0.15\\
Oph     &       (355.81,16.63)  &       5       &       (34.94,16.63)   &  11.3& 0.56  &  12.1& 2.80\\
Perseus &       (159.31,-20.25) &       4       &       (148.44,-19.88) &  4.72& 0.29  &  4.82& 0.76\\
Taurus  &       (173.19,-14.75) &       6       &       (143.94,-19.50) &  16.4& 6.69  &  15.4& 1.27\\
Orion A &       (212.19,-19.13) &       4       &       (233.69,-19.13) &  9.35& 4.06  &  9.10& 9.34\\
Orion B &       (204.56,-13.75) &       4       &       (233.69,-19.13)  &  9.26& 10.2  &  8.70& 14.3\\
Chamaeleon      &       (300.43,-16.13) &       5.5     &       (283.81,-16.13) &  4.56& 1.21  &  4.46& 3.53\\
Mon R2  &       (213.81,-12.63) &       1.5     &       (233.81,-18.75) &  1.91& 1.17  &  1.81& 9.47\\
\hline
\hline
RCrA    &       (0.56,-19.63)   &       3       &       (6.94,-19.63)   &  0.73& 1.80  &  0.33& 0.06\\
Oph     &       (355.81,16.63)  &       5       &       (34.94,16.63)   &  10.2& 0.51  &  5.46& 1.27\\
Perseus &       (159.31,-20.25) &       4       &       (148.44,-19.88) &  4.02& 0.24  &  2.01& 0.32\\
Taurus  &       (173.19,-14.75) &       6       &       (143.94,-19.50) &  13.6& 5.56  &  6.32& 0.52\\
Orion A &       (212.19,-19.13) &       4       &       (233.69,-19.13) &  7.75& 3.37  &  3.73& 3.83\\
Orion B &       (204.56,-13.75) &       4       &       (233.69,-19.13)  &  7.56& 8.32  &  3.55& 5.83\\
Chamaeleon      &       (300.43,-16.13) &       5.5     &       (283.81,-16.13) &  3.78& 1.00  &  1.86& 1.47\\
Mon R2  &       (213.81,-12.63) &       1.5     &       (233.81,-18.75) &  1.56& 0.95  &  0.75& 3.93\\
\hline
\end{tabular}
\tablefoot{The second and third columns are the position and size of the source region. The forth column is the position of the background estimation region, whose size is the same as the source region. The fluxes and errors above the double horizontal lines in 0.1-100 GeV are in units of $\rm 10^{-7} \ ph \ cm^{-2} \ s^{-1} $ and $\rm 10^{-9} \ ph \ cm^{-2} \ s^{-1} $, respectively.
The fluxes and errors above the double horizontal lines in 1-500 GeV are in units of $\rm 10^{-8} \ ph \ cm^{-2} \ s^{-1} $ and $\rm 10^{-10} \ ph \ cm^{-2} \ s^{-1} $, respectively.
The fluxes and errors below the double horizontal lines are in units of $\rm 10^{-10} \ erg \ cm^{-2} \ s^{-1} $ and $\rm 10^{-12} \ erg \ cm^{-2} \ s^{-1} $, respectively.}

\label{tableinfofermi}
\end{table*}

\begin{table*}
\centering
\caption{Change in gamma-ray flux of the first three molecular clouds for different background models.}
\begin{tabular}{lcccccccccccccccc}
\hline
\hline
Name & 1 & 2 &3 &4 &5 &6 &7 &8 &9 &10 &11& 12& 13& 14 & 15 & 16 \\
\hline
RCrA & 0.13 & 0.06 & 0.15 & 0.07 & 0.11 &  0.03 & 0.14 & 0.04 & 0.11 &  0.03 &  0.13 & 0.04 &0.09 & -0.01 & 0.12  &0.01 \\

Oph     & 0.05& 0.05 & 0.05 & 0.06 & 0.05& 0.06 & 0.06 & 0.07 & 0.05 & 0.06&0.06& 0.07&
 0.05 & 0.08 & 0.07 & 0.08 \\
Perseus & 0.07 &0.02 & 0.08 &0.05 &0.07 &0.06& 0.09 & 0.05 & 0.06 & 0.02 & 0.08& 0.04 & 0.07 & 0.02 & 0.08 & 0.05\\
\hline
\end{tabular}
\tablefoot{The values represent the ratio between $\rm (F2 - F1)$ and $\rm F1$, where $\rm F1 $ is derived  using the method in the second paragraph of section 2, and $\rm F2$ is derived using the method in the third paragraph of section 2.\\
1: (Lorimer, 4 kpc, 150 K, and 2 mag); 2: (Lorimer, 4 kpc, 150 K, and 5 mag); 3: (Lorimer, 4 kpc, $10^5$ K, and 2 mag); 4: (Lorimer, 4 kpc, $10^5$ K, and 5 mag);
5: (Lorimer, 10 kpc, 150 K, and 2 mag); 6: (Lorimer, 10 kpc, 150 K, and 5 mag); 7: (Lorimer, 10 kpc, $10^5$ K, and 2 mag); 8: (Lorimer, 10 kpc, $10^5$ K, and 5 mag);
9: (SNR, 4 kpc, 150 K, and 2 mag); 10: (SNR, 4 kpc, 150 K, and 5 mag); 11: (SNR, 4 kpc, $10^5$ K, and 2 mag); 12: (SNR, 4 kpc, $10^5$ K, and 5 mag);
13: (SNR, 10 kpc, 150 K, and 2 mag); 14: (SNR, 10 kpc, 150 K, and 5 mag); 15: (SNR, 10 kpc, $10^5$ K, and 2 mag); 16: (SNR, 10 kpc, $10^5$ K, and 5 mag).
}

\label{tablerates}
\end{table*}

\begin{table*}
\centering
\caption{Parameters and gamma-ray luminosities of star-forming galaxies. }
\begin{tabular}{lccccc}
\hline
\hline
Name & Distance    & $ L_{\rm 0.1-100 \ GeV}$ & $ L_{\rm 1-500 \ GeV}$ & SFR & $M_{\odot,9}$\\
 & [Mpc]    & [$ \rm erg \ s^{-1}$] & [$ \rm erg \ s^{-1}$]  & [$M_{\odot} \ \rm yr^{-1}$] & \\
\hline
Milky Way       &       ...     &       $(8.20\pm 2.40)\times10^{38}$   &  ... & 1-3\tablefootmark{1} & $4.90\pm 0.45$\tablefootmark{2}\\
LMC     &       0.05    &       $(4.70\pm0.50)\times10^{37}$    & $(2.94\pm0.05)\times10^{37}$ & 0.20-0.25\tablefootmark{3} & $0.53\pm 0.02$\tablefootmark{4}\\
SMC     &       0.06    &       $(1.10\pm0.30)\times10^{37}$    & $(7.57\pm0.34)\times10^{36}$ & 0.04-0.08\tablefootmark{5}  & $0.45\pm 0.04$\tablefootmark{6} \\
M31     &       0.78    &       $(4.60\pm1.00)\times10^{38}$    & $(6.31\pm0.21)\times10^{37}$  &0.35-1\tablefootmark{1}& $7.66\pm 2.21$\tablefootmark{7}\\
NGC 253 &       2.5     &       $(6.00\pm2.00)\times10^{39}$    & $(3.51\pm0.40)\times10^{39}$ & 3.5-10.4\tablefootmark{8} & $2.20$\tablefootmark{9} \\
M82     &       3.4     &       $(1.50\pm0.30)\times10^{40}$    & $(7.74\pm0.28)\times10^{39}$  & 13-33\tablefootmark{10} & $2.72$\tablefootmark{9} \\
NGC 2146        &       15.2    &       $(4.62\pm2.43)\times10^{40}$    & $(3.39\pm1.20)\times10^{40}$ & 26.6-79.7\tablefootmark{11} & $5.98$\tablefootmark{9}\\
Arp 220 &       74.7    &       $(1.78\pm0.30)\times10^{42}$    & $(6.24\pm2.21)\times10^{41}$ & 254.8-764.3\tablefootmark{11} & $37.5$\tablefootmark{9} \\
\hline
\end{tabular}
\tablefoot{The distances are provided by \citet{2012ApJ...755..164A}. The gamma-ray luminosities of the Milky Way $L_{0.1-100 \ \rm GeV}$ have been estimated using a numerical model of CR propagation and interactions in the ISM \citep{2010ApJ...722L..58S}. $M_{\odot,9}$ is $M_{\odot}/10^{9}$.}\\
\textbf{References.} \tablefoottext{1}{\citet{2009A&A...505..497Y}}; \tablefoottext{2}{\citet{2007A&A...465..839P}}; \tablefoottext{3}{\citet{2007MNRAS.382..543H}}; \tablefoottext{4}{\citet{2003MNRAS.339...87S,2008ApJS..178...56F}};
\tablefoottext{5}{\citet{2004A&A...414...69W}}; \tablefoottext{6}{\citet{1999MNRAS.302..417S,2007ApJ...658.1027L}}; \tablefoottext{7}{\citet{2009ApJ...695..937B,2006A&A...453..459N}}; \tablefoottext{8}{\citet{2006AJ....132.1333L}}; \tablefoottext{9}{\citet{2004ApJ...606..271G}}; \tablefoottext{10}{\citet{2003ApJ...599..193F}}; \tablefoottext{11}{\citet{2005ApJ...621..139C}}.
\label{tableinfoGax}
\end{table*}

\begin{table*}
\centering
\caption{Results of the two-parameter correlation for $L_{1-500 \ \rm GeV}$ and SFR in our sample.}
\begin{tabular}{cccccc}
\hline
\hline
$r$    & $p$ & $a$ &$b$ & $\sigma$ & $\delta$ \\
\hline
\multicolumn{6}{c}{Clouds}  \\
\hline
0.73&0.004 & $37.31^{+1.95}_{-1.92}$ & $1.02^{+0.49}_{-0.47}$ & $0.55^{+0.25}_{-0.15}$ & 0.41\\
\hline
\multicolumn{6}{c}{Galaxies}  \\
\hline
0.996 & $\sim 10^{-6}$ & $38.29^{+0.09}_{-0.10}$ & $1.28^{+0.07}_{-0.07}$ & $0.21^{+0.11}_{-0.06}$ & 0.15\\
\hline
\end{tabular}
\tablefoot{$r$ is the Pearson correlation coefficient, and $p$ is the chance probability. We model the two-parameter correlation using the form ${\rm log}L_{\gamma} = a + b{\rm log(SFR)}$. $\sigma $ is introduced to accommodate intrinsic scatter and measurement errors. $\delta$ is the dispersion of a regression model with standard deviation of $L_{\gamma}^{\rm r}$ from $L_{\gamma}$, where $r$ marks the $L_{\gamma}$ value derived from the regression model. The coefficients of $a, \ b,$ and $\sigma$ are constrained simultaneously by maximizing the joint likelihood function.}
\label{tableLumSFR}
\end{table*}

\begin{table*}
\centering
\caption{Results of the two-parameter correlation for $L_{1-500 \ \rm GeV}$ and M in our sample.}
\begin{tabular}{cccccc}
\hline
\hline
$r$    & $p$ & $a$ &$b$ & $\sigma$ & $\delta$\\
\hline
\multicolumn{6}{c}{Clouds-dense}  \\
\hline
0.85    &       0.007 & $30.23^{+0.89}_{-0.90}$ & $0.92^{+0.28}_{-0.28}$ & $0.42^{+0.18}_{-0.11}$ & 0.31\\
\hline
\multicolumn{6}{c}{Clouds-total}  \\
\hline
0.97    &       $\sim 10^{-4}$ & $28.90^{+0.54}_{-0.55}$ & $1.02^{+0.14}_{-0.14}$ & $0.19^{+0.09}_{-0.06}$ & 0.15\\
\hline
\multicolumn{6}{c}{Galaxies}  \\
\hline
0.80 & 0.03 & $18.86^{+8.59}_{-8.37}$ & $2.14^{+0.88}_{-0.90}$ & $1.43^{+0.77}_{-0.41}$ & 1.00\\
\hline
\end{tabular}
\tablefoot{$r$ is the Pearson correlation coefficient, and $p$ is the chance probability. We model the two-parameter correlation using the form ${\rm log }L_{\gamma} = a + b{\rm log}(M)$. $\sigma $ is introduced to accommodate intrinsic scatter and measurement errors. $\delta$ is the dispersion of a regression model with standard deviation of $L_{\gamma}^{\rm r}$ from $L_{\gamma}$, where $r$ marks the $L_{\gamma}$ value derived from the regression model. The coefficients of $a, \ b,$ and $\sigma$ are constrained simultaneously by maximizing the joint likelihood function.}
\label{tableLumM}
\end{table*}

\begin{table*}
\centering
\caption{Results of the three-parameter correlation for $L_{1-500 \ \rm GeV}$, SFR, and M in our sample.}
\begin{tabular}{ccccccc}
\hline
\hline
$r$    & $p$ & $a$ &$b$ & $c$ & $\sigma$ & $\delta$\\
\hline
\multicolumn{7}{c}{Clouds}  \\
\hline
0.979   &       $\sim 10^{-5}$ & $26.67^{+2.02}_{-1.90}$ & $-0.33^{+0.29}_{-0.29}$ & $1.23^{+0.22}_{-0.22}$ & $0.19^{+0.10}_{-0.06}$&0.12\\
\hline
\multicolumn{7}{c}{Galaxies}  \\
\hline
0.997   &       $\sim 10^{-6}$ & $40.49^{+2.24}_{-2.25}$ & $1.37^{+0.12}_{-0.12}$ & $-0.24^{+0.25}_{-0.25}$& $0.21^{+0.14}_{-0.07}$       & 0.12\\
\hline
\end{tabular}
\tablefoot{$r$ is the Pearson correlation coefficient, and $p$ is the chance probability. We model the three-parameter correlation using the form ${\rm log}L_{\gamma} = a + b{\rm log(SFR)} + c{\rm log}(M) $. $\sigma $ is introduced to accommodate intrinsic scatter and measurement errors. $\delta$ is the dispersion of a regression model with standard deviation of $L_{\gamma}^{\rm r}$ from $L_{\gamma}$, where $r$ marks the $L_{\gamma}$ value derived from the regression model. The coefficients of $a, \ b,\ c,$ and $\sigma$ are constrained simultaneously by maximizing the joint likelihood function.}
\label{tableGeV}
\end{table*}

\begin{table*}
\centering
\caption{Results of the three-parameter correlation for $L_{0.1-100 \ \rm GeV}$, SFR, and M in our sample.}
\begin{tabular}{ccccccc}
\hline
\hline
$r$    & $p$ & $a$ &$b$ & $c$ & $\sigma$ & $\delta$\\
\hline
\multicolumn{7}{c}{Clouds}  \\
\hline
0.976   &       $\sim 10^{-5}$ & $27.16^{+1.98}_{-2.13}$ & $-0.32^{+0.30}_{-0.32}$ & $1.22^{+0.24}_{-0.23}$ & $0.20^{+0.10}_{-0.06}$ & 0.13\\
\hline
\multicolumn{7}{c}{Galaxies}  \\
\hline
0.998   &       $\sim 10^{-8}$ & $35.70^{+1.61}_{-1.59}$ & $1.16^{+0.09}_{-0.09}$ & $0.31^{+0.18}_{-0.18}$& $0.16^{+0.09}_{-0.05}$        & 0.11\\
\hline
\end{tabular}
\tablefoot{$r$ is the Pearson correlation coefficient, and $p$ is the chance probability. We model the three-parameter correlation using the form ${\rm log}L_{\gamma} = a + b{\rm log(SFR)} + c{\rm log}(M) $. $\sigma $ is introduced to accommodate intrinsic scatter and measurement errors. $\delta$ is the dispersion of a regression model with standard deviation of $L_{\gamma}^{\rm r}$ from $L_{\gamma}$, where $r$ marks the $L_{\gamma}$ value derived from the regression model. The coefficients of $a, \ b,\ c,$ and $\sigma$ are constrained simultaneously by maximizing the joint likelihood function.}
\label{tableMeV}
\end{table*}

\clearpage

\end{document}